\pdfoutput=1
\documentclass[11pt]{article}

\usepackage[preprint]{acl}
\usepackage{threeparttable}
\usepackage{times}
\usepackage{latexsym}
\usepackage{booktabs} % 引入 booktabs 宏包
\usepackage{multirow}
\usepackage[T1]{fontenc}
\usepackage{graphicx}
\usepackage{balance}
\usepackage{nopageno}
\usepackage[utf8]{inputenc}

\usepackage{microtype}

\usepackage{inconsolata}

\usepackage{graphicx}

\title{Breaking the Hourglass Phenomenon of Residual Quantization: Enhancing the Upper Bound of Generative Retrieval}

\author{Zhirui Kuai\textsuperscript{\rm 1,\textdagger} \and Zuxu Chen\textsuperscript{\rm 2,\textdagger},  \and Huimu Wang\textsuperscript{\rm 3,\thanks{Corresponding Author. \textdagger Equal Contribution.}} and Mingming Li\textsuperscript{\rm 3,*} \\
   \bf{Dadong Miao\textsuperscript{\rm 3} \and Binbin Wang\textsuperscript{\rm 3} \and Xusong Chen\textsuperscript{\rm 3} \and Li Kuang\textsuperscript{\rm 1} \and Yuxing Han\textsuperscript{\rm 2,*} }\\ 
 \bf{Jiaxing Wang\textsuperscript{\rm 3} \and Guoyu Tang\textsuperscript{\rm 3}  \and Lin Liu\textsuperscript{\rm 3}  \and Songlin Wang\textsuperscript{\rm 3} \and  Jingwei Zhuo\textsuperscript{\rm 3}} \\
 \textsuperscript{\rm 1}Central South University, School of Computer Science and Engineering, China \\
   \textsuperscript{\rm 2}Shenzhen International Graduate School, Tsinghua University, China \\ 
\textsuperscript{\rm 3}JD.com, Beijing, China \\
{\fontsize{11.2}{10}\selectfont kuaizhirui@csu.edu.cn  \and    {chen-zx22@mails.tsinghua.edu.cn} \and {yuxinghan@sz.tsinghua.edu.cn}} \\
   {\fontsize{11.2}{10}\selectfont \{wanghuimu1,limingming65,zhuojingwei\}@jd.com}}

\begin{document}
\maketitle
\begin{abstract}
\vspace{-0.2cm}
Generative retrieval (GR) has emerged as a transformative paradigm in search and recommender systems, leveraging numeric-based identifier representations to enhance efficiency and generalization. Notably, methods like TIGER employing Residual Quantization-based Semantic Identifiers (RQ-SID), have shown significant promise in e-commerce scenarios by effectively managing item IDs. However, a critical issue termed the "\textbf{Hourglass}" phenomenon, occurs in RQ-SID, where intermediate codebook tokens become overly concentrated, hindering the full utilization of generative retrieval methods. This paper analyses and addresses this problem by identifying path sparsity and long-tailed distribution as the primary causes. Through comprehensive experiments and detailed ablation studies, we analyze the impact of these factors on codebook utilization and data distribution. Our findings reveal that the "Hourglass" phenomenon substantially impacts the performance of RQ-SID in generative retrieval. We propose effective solutions to mitigate this issue, thereby significantly enhancing the effectiveness of generative retrieval in real-world E-commerce applications.
\vspace{-0.3cm}
\end{abstract}
\vspace{-0.3cm}
\section{Introduction}
% \vspace{-0.3cm}
In recent years, GR has surfaced as a groundbreaking retrieval paradigm, marking significant advancements in search and recommendation environments including recommender systems \cite{tiger,tan2024towards,wang2024enhanced}, search question answering \cite{liu2023webglm,qin2023webcpm}, and E-commerce retrieval \cite{DSI,NCI,li2024generative}. In this paradigm, target items are initially represented as identifiers (e.g., numbers, subwords, n-grams, token IDs, URLs, semantic codes). Subsequently, leveraging input information such as queries and user details, large models are employed to output the final items in an end-to-end manner. This approach not only enhances retrieval efficiency but also improves the model’s generalization capability.

In generative retrieval, numeric-based identifier representation methods are widely adopted in the industry due to their simplicity, efficiency, and strong generalization, especially in long behavior sequence recommendations. These methods significantly reduce sequence lengths and accelerate the inference process. Notable methods include DSI \cite{DSI}, NCI \cite{NCI}, TIGER \cite{tiger}, GDR \cite{yuan2024generative}, and GenRet \cite{sun2024learning}. Among these, the TIGER method generates Semantic Identifiers (SID) through Residual Quantization (RQ) \cite{lee2022autoregressive,zeghidour2021soundstream}, effectively capturing both semantic information and hierarchical structures. This approach is particularly advantageous in item-dominated e-commerce scenarios, where it accurately reflects the complex hierarchical relationships and semantic features inherent in e-commerce data, thereby significantly enhancing recommendation performance.
\begin{figure}
    \vspace{0.51 cm}
    \centering
    \includegraphics[width=1\linewidth]{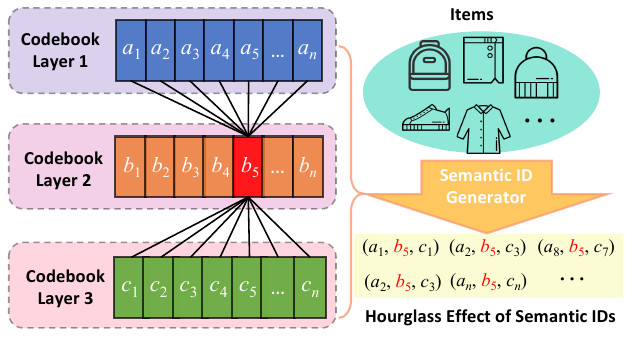}
    \caption{The Hourglass Phenomenon of Semantic IDs}
    \label{fig:IntroFig}
    % \vspace{-0.3cm}
\end{figure}

It is important to highlight that the performance upper bound of RQ-based methods critically depends on the generation of SID. However, we have identified a significant "hourglass" phenomenon in SID produced via RQ, as illustrated in Figure \ref{fig:IntroFig}. Specifically, the codebook tokens in the intermediate layers are excessively concentrated, leading to a one-to-many and many-to-one mapping structure. This concentration results in path sparsity, where the matching paths for the item constitute a minimal fraction of the total path space and a long-tail distribution of intermediate layer tokens with a majority of SID concentrated in a few head tokens. This hourglass effect is particularly exacerbated in datasets with long-tail characteristics, which substantially constrains the representational capacity of GR methods. The underlying cause of this issue stems from the intrinsic nature of progressively quantizing high-dimensional vector residuals.

Furthermore, we analyzed the process of generating SID from residuals, demonstrating that sparsity and long-tail distributions are inevitable. To assess the general impact of SID on downstream GR tasks, we trained models of different scales (such as 0.8B, 7B) and types (Qwen1.5 \cite{bai2020sparterm}, Baichuan2 \cite{yang2023baichuan}, LLaMA2 \cite{touvron2023llama}) based on RQ-SID. Through a series of experiments, including altering the distribution of Semantic IDs by interacting with the first and second layers and swapping tokens between the first and second layers, we not only confirmed the existence of the Hourglass effect but also detailed its specific impact on model performance. This analysis provides a robust foundation for future model optimization.

To alleviate the hourglass effect, we propose two straightforward yet effective methods: the heuristic approach and the adaptive variable-length token strategy. The heuristic method involves directly removing the second layer, while curtailing the long-tail impact, it may lead to insufficient spatial capacity. The second method implements an adaptive token distribution adjustment to remove the top tokens from the second layer, thereby transforming the semantic ID into a variable-length structure. This strategy ensures that the overall distribution remains consistent while effectively mitigating the hourglass effect by selectively token removal. Extensive experimental results reveal that although both methods are straightforward, they successfully alleviate the impact of the hourglass effect to varying extents. Notably, the adaptive variable-length token strategy method emerges as the most effective.
% In our research, through detailed comparative experiments and a series of ablation studies (such as codebook utilization and data distribution analysis), it is demonstrated that data sparsity and long-tail characteristics are the primary causes of the "hourglass" phenomenon.

The contributions of this paper can be summarized as follows:
\begin{itemize}
    \item  To our knowledge, this is the first study to systematically investigate the deficiencies of residual quantization-based semantic identifiers in generative retrieval, specifically identifying the "hourglass" phenomenon where intermediate layer codebook tokens are overly concentrated.
    % To the best of our knowledge, this is the first study that systematically explores the deficiencies of residual quantization-based semantic identifiers (RQ-SID) in generative retrieval (GR). By identifying the "hourglass" phenomenon, where tokens in intermediate layer codebooks are overly concentrated, we highlight a critical issue that has been overlooked in existing research.
    \item  We conduct thorough experiments and ablation studies that reveal path sparsity and long-tail distributions as the primary causes of the "hourglass" effect, limiting the representation and performance capabilities of generative models.
    
    % Through comprehensive experiments and detailed ablation studies, we thoroughly analyze and demonstrate the primary causes of the deficiencies in RQ-SID. Our findings reveal that data sparsity and long-tail characteristics are the main contributors to the "hourglass" phenomenon, which limits the representation capabilities and overall performance of GR models.
    
    \item  We propose and validate a novel method to alleviate the \textbf{"hourglass"} effect, which significantly enhances model performance by improving codebook utilization and addressing token long-tail distributions.
    
    % We propose a series of innovative improvement methods aimed at mitigating the "hourglass" effect and significantly enhancing the model's capabilities. These methods not only improve codebook utilization but also address the long-tail distribution of tokens, thereby enhancing the effectiveness and efficiency of generative retrieval in real-world scenarios.
\end{itemize}

\section{Related Works}
Recent advancements in generative retrieval have significantly influenced various domains, such as recommendation systems, search question answering, and E-commerce retrieval. This paradigm shift, as evidenced by works like \cite{DSI,NCI,wang2024enhanced,li2024generative}, involves representing target items using identifiers such as numbers, sub-words, and semantic codes. 

Within the industry, numeric-based identifier representation methods are prevalent due to their simplicity and efficiency. These methods, including DSI \cite{DSI}, NCI \cite{NCI}, TIGER \cite{tiger}, GDR \cite{yuan2024generative}, and GenRet \cite{GenRet-baidu}, are particularly effective in long behavior sequence recommendations. They reduce sequence lengths and accelerate inference processes. Notably, the TIGER method employs RQ\cite{lee2022autoregressive,zeghidour2021soundstream} to generate SID, capturing semantic information and hierarchical structures. This is especially beneficial in item-dominated e-commerce contexts, where complex hierarchical relationships and semantic features are crucial for enhancing recommendation performance. However, the performance upper limit of RQ-based methods largely depends on the generation of SID, which is also the central focus of analysis and discussion in this paper.

\section{Preliminary}

% a novel retrieval paradigm, generative retrieval, has achieved significant breakthroughs in search and recommendation contexts, such as recommendation recall, search Q&A, and e-commerce search, among others. This approach initially represents the target text or item as an identifier (e.g., digits, sub-words, n-grams, token IDs, URLs, semanticID), and then, given input information, such as a query or user information, the large model directly generates the final item in an end-to-end manner. Compared to the traditional k-NN dual-tower models, this has a notable advantage in interaction while also reducing the quantization loss and index storage overhead associated with two-stage processes. Due to the simplicity, efficiency, and good generalizability of code-based identifier representation methods, especially in the domain of long-sequence recommendation, it can significantly reduce sequence length and accelerate the inference process. Therefore, this discussion focuses on semanticID-based inference methods.

\subsection{Residual Quantization}
Residual-quantized is a multi-level vector quantizer that applies quantization on residuals to generate a tuple of codewords (i.e., Semantic IDs). Residual-quantized variational AutoEncoder (RQ-VAE) \cite{tiger,lee2022autoregressive,zeghidour2021soundstream} is jointly trained by updating the quantization codebook and the encoder-decoder reconstruction parameters. 

Support that there is a vector $\textbf{x} \in \mathcal{R}^D$, we aim to quantize it using  $L$ codebooks ($L$ layer) of $M$ elements each, where codebook could be denoted as $\textbf{C} \in \mathcal{R}^{L \times M \times D}$, $D$ is the dimension of vector.
When $l=1$, the initial residual is simply defined as $\textbf{r}_1=\textbf{x} $. 
Then, $\textbf{r}_l$ is quantized by mapping it to the nearest embedding from that layer’s codebook $\textbf{C}_l \in \mathcal{R}^{M \times D}$. The index of the closest embedding at this layer could be computed as follows:
 \begin{equation}
     c_l = \arg \min_{m \in M} \parallel \textbf{r}_l - \textbf{C}_{l,m} \parallel_{2}^{2}
 \end{equation}
where $c_l$ represents the $l$-th codeword(semantic ID).
Note that, at the $l$-th layer ($l$ > 1), the residual is: 
% $\textbf{r}_l=\textbf{r}_{l-1} - \textbf{C}_{l,c_{l-1}}$.
 \begin{equation}
    \textbf{r}_l=\textbf{r}_{l-1} - \textbf{C}_{l,c_{l-1}}
\end{equation}
The above process is repeated recursively $L$ times to get a tuple of $L$ codewords that represent the Semantic ID for the given $\textbf{x}$, denoted as ($c_1, c_2, \dots,c_L$).

% For the other layer $l>1$, the residual is defined as  
% \begin{equation}
%     \textbf{r}_l=\textbf{r}_{l-1} - \textbf{C}_{l,c_{l-1}}
% \end{equation}
% Then, similar to the 1-th layer, the code for this level is computed
% by finding the embedding in the codebook $\textbf{C}_l \in \mathcal{R}^{M \times D}$ for the current layer which is nearest to $\textbf{r}_l$. 

To reconstruct the raw vector, we sum the corresponding codebook elements as:
\begin{equation}
    {\hat{\textbf{x}}} = \sum_{l=0}^{L} \textbf{C}_{l,c_{l}}
\end{equation}

This method could approximate the raw vector from a coarse-to-fine granularity by the norm of residuals decreasing, i.e., $\Vert \hat{\textbf{x}} - {\textbf{x}}\Vert^2 < \epsilon$,  $\epsilon \ll 0.001$.

\subsection{Generative Retrieval}
Generative retrieval \cite{NCI,DSI,se-DSI-tang2023semantic,seal-bevilacqua2022autoregressive,ppo-zhou2023enhancing}, has been proposed in the recommendation field, search field and question-answer field.
These models advocate generating identifiers of target passages/items directly through the autoregressive language models. 
% Existing work could be divided into two categories based on identifier types:
% 1) Numeric-based \cite{NCI,zhuang2022bridging,rajput2024recommender,xiaohongshu-semanticid-yuan2024generative}, they assign numeric identifiers in various ways, e.g., atomic, naive, and semantic. 
% 2) lexical identifier-based methods \cite{seal-bevilacqua2022autoregressive,glen-lee2023glen,li2023multiview} using the n-grams, title, and URLs as the document identifiers. They could leverage the knowledge of PLMs to decode identifiers, exploring the benefit of pre-trained vocabulary space.
% The lexical identifier-based methods show potential in terms of interpretability and generalization capabilities, especially in the era of large language models. Thus, we continue to explore along these lines of methods in this paper.

% In the field of E-commerce, there exist several crucial challenges. 
% Firstly, the task of query2title generation poses difficulties. Specifically, product titles tend to be lengthy on average, whereas user-entered query words are typically short. Attempting to directly generate lengthy titles can result in significant hallucination issues. While some efforts have been made to utilize pre-trained semantic IDs as document identifiers \cite{NCI,DSI,xiaohongshu-semanticid-yuan2024generative} to simplify the task into query-to-semanticID and reduce complexity, this approach heavily relies on external document representations, deviating significantly from the language itself and necessitating additional calibration, thereby diminishing result interpretability.

In personalized search scenarios, a core task is to provide the most relevant candidates that the user is likely to purchase based on their given query and historical interaction behaviors. In this paper, we re-frame this task as a Next Token Prediction (NTP) problem utilizing LLM and Semantic ID. Specifically, given user $u$, query $q$, and  the user's historical item sequence,  we first convert the sequence into a Semantic ID sequence, denoted as $Seq :=\left\{\underbrace{\left(c_{1,1},\cdot,c_{1,M}\right)}_{item_1}; \underbrace{\left(c_{2,1},\cdot,c_{2,M}\right)}_{item_2};\ldots; \underbrace{\left(c_{t,1},\cdot,c_{t,M}\right)}_{item_t}  \right\}$
where ($c_{i,1}$, $\cdot$, $c_{i,M}$) denotes the $M$-length Semantic ID for $item_i$. The LLM is then trained to predict the Semantic ID of $item_{t+1}$, represented as ($c_{t+1,1}$, $\cdot$, $c_{t+1,M}$). 
The generation objective could be formulated as,
\begin{equation}
   \mathcal{L}_{sft} = - \sum_i^{M} \log {p_\theta}(i|q,u, Seq, I_{<i}) 
\end{equation}
where $I_{<i} = \{c_{t+1,1},\cdots,c_{t+1,i}\}$,  $p_\theta $ is the supervised fine-tuning (SFT) model.

% In recommendation scenarios, a common task is to predict the next item a user is likely to purchase based on their historical interaction behavior. 
% In this paper, we re-frame this task as a Next Token Prediction (NTP) problem utilizing LLM and Semantic ID. Specifically, we convert the user's historical item sequence into a Semantic ID sequence:
% ($c_{1,0}$,$\ldots$,$c_{1,m-1}$,$c_{2,0}$,$\ldots$,$c_{2,m-1}$,$\ldots$,$c_{n,0}$,$\ldots$,$c_{n,m-1}$)
% where ($c_{i,0}$, $\ldots$, $c_{i,m-1}$) denotes the $m$-length Semantic ID for $item_i$. The LLM is then trained to predict the Semantic ID of $item_{n+1}$, represented as ($c_{n+1,0}$, $\ldots$, $c_{n+1,m-1}$). 

% Given the generative nature of LLM, it is possible that a generated Semantic ID from the decoder does not match an item in the recommendation corpus. However, there are commonly used methods like the FM-index \cite{ferragina2000opportunistic} and Tire \cite{bodon2003trie} that can mitigate this issue, and thus this paper will not elaborate on them further.

\section{Problem of GR based on RQ}

\subsection{Hourglass Phenomenon}

To generate the semantic IDs used RQ, we first leverage the query-item data from billions of search logs within the company to train dual-tower models such as DSSM and BERT \cite{li2020symmetric,fan2019mobius,karpukhin2020dense,mixpq,qiu2022pre}. Subsequently, we obtain the embeddings for hundreds of millions of items using the item tower. Finally, we employ RQ to generate semantic IDs for all items.

\begin{figure}[t]
    \centering
    \includegraphics[scale=0.12]{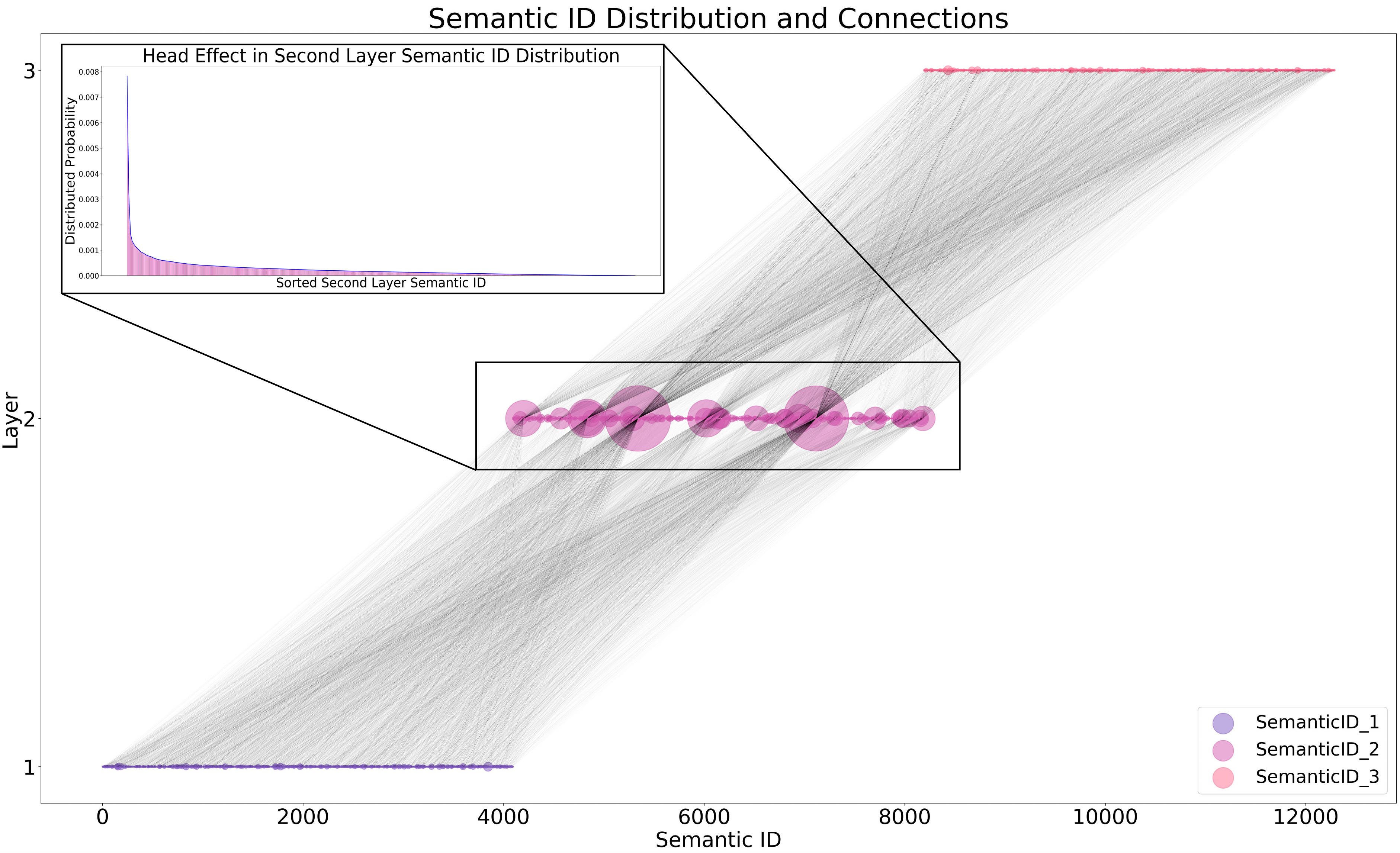}
    \caption{Distribution and Connections of Semantic IDs}
    \label{fig:three-layer}
    % \vspace{-0.4cm}
\end{figure}

Upon the successful generation of semantic IDs, we proceed to aggregate and compute the three-layer distribution maps for all items. As illustrated in Figure \ref{fig:three-layer}, it is evident that the second layer of the Semantic ID architecture is concentrated with a substantial number of routing nodes. The overall distribution of the three-layer code exhibits an hourglass phenomenon.  

To investigate the generalizability of this phenomenon, we conducted multiple visualization experiments under various parameter combinations, e.g., code table size and number of layers. As shown in Figure \ref{fig:dis-label} in the appendix, the results indicate that the hourglass effect is highly pronounced, and the path distribution among the tokens across the three layers of the code table is relatively sparse.

\begin{figure*}[t]
    \centering
    \includegraphics[width=1.0\linewidth]{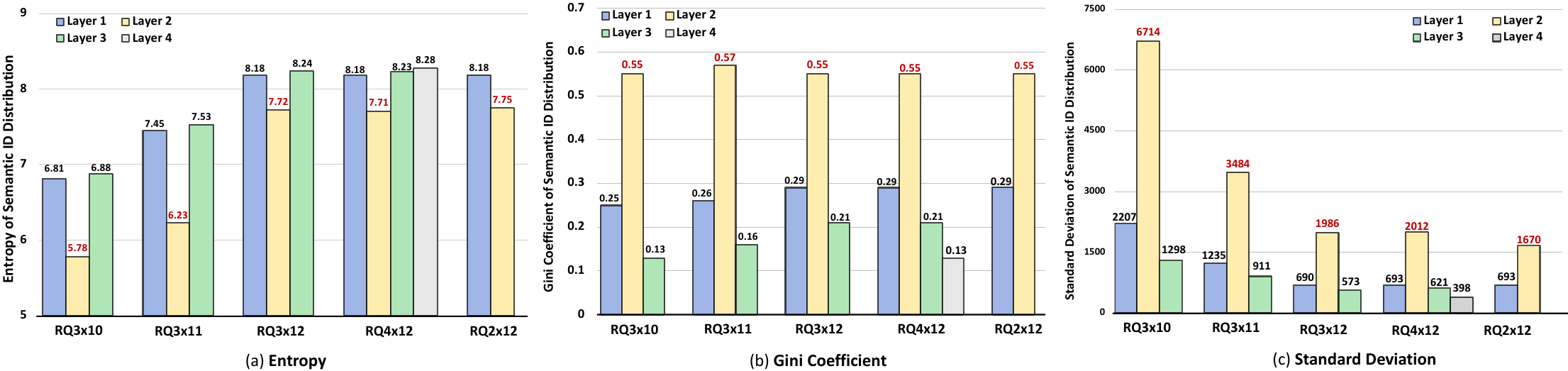}
    \caption{Illustrating the Hourglass Phenomenon in Semantic IDs with Different Statistical Metrics}
    \label{fig:tokendis}
\end{figure*}

% From the Figure, we can see that there is small entropy, a high Gini coefficient, and a large standard deviation (SD)  of the token distribution in the second layer indicating that the distribution is highly skewed and has a long tail.  
Additionally, based on the aforementioned experiments, we conducted statistical analyses of the token distribution in the second layer using three metrics: entropy \cite{shannon1948mathematical}, Gini coefficient \cite{yitzhaki1979relative}, and standard deviation \cite{pal2019introduction}, as shown in the Figure \ref{fig:tokendis}. The results indicate that the token distribution in the second layer exhibits low entropy, high Gini coefficient, and large standard deviation, suggesting that the distribution is highly skewed and exhibits a long-tail effect.

Overall, this hourglass phenomenon is statistically evidenced in the code table by path sparsity and a long-tail distribution of tokens. 1) Path sparsity, resulting from the Semantic ID structure, leads to low code table utilization. 2) The long-tail distribution indicates that in the intermediate layer, a predominant number of routes converge on a single token.

% This phenomenon is statistically manifested in the code table as path sparsity and token long-tail distribution. The former refers to the sparsity of paths formed by the Semantic ID, resulting in low utilization of the code table. The latter indicates that in the distribution of tokens in the intermediate layer, the vast majority of routes converge on a single token.

% It is obvious that the overall architecture displays an hourglass effect, with significant routing nodes concentrated in the second layer, resulting in sparse pathways(low space utilization), and a pronounced long-tail impact for each layer code.

\subsection{Analysis of Residual Quantization}
% According to the above analysis of distribution of Semantic IDs, we find that this hourglass phenomenon is ubiquitous on different RQ parameters.
\begin{figure*}[t]
    \centering
    \includegraphics[width=1\linewidth]{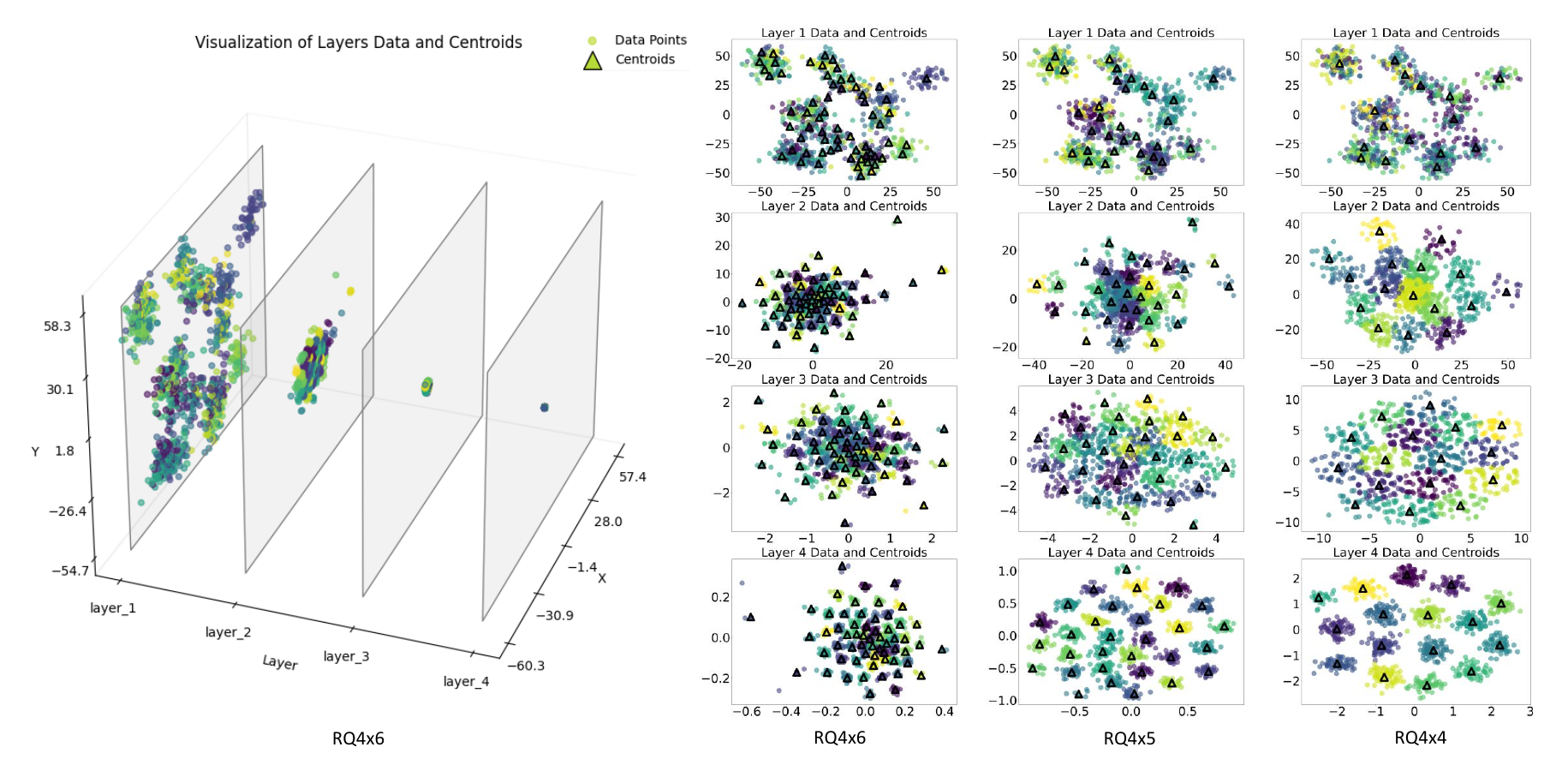}
    \caption{Hierarchical Residual Reduction and Dimensional Analysis Across Layers}
    \label{fig:Analysis}
    % \vspace{-1em}
\end{figure*}
% Based on the hourglass phenomenon demonstrated in the aforementioned experiments, along with the corresponding path sparsity and long-tail distribution of tokens, it is imperative to conduct an in-depth analysis to understand the underlying causes of this phenomenon.
To explore the causes of the hourglass phenomenon, we will conduct an in-depth analysis and discussion based on the operating mechanism of the RQ.
Without loss of generality, we consider two distributions of raw embedding:  un-uniform and uniform, denoted as $\textbf{X} = \left\{ \textbf{x}|\textbf{x} \in \textbf{X} \right\} \in \mathcal{R}^{N \times M}$, $N$ is the size of the dataset.
Now, we use the RQ to produce the semantic ID for $\textbf{X}$.  

% \begin{figure*}[t]
%     \centering
%     \includegraphics[width=1\linewidth]{figs/RQ4x6.pdf}
%     \caption{Hierarchical Residual Reduction and Dimensional Analysis Across Layers}
%     \label{fig:Analysis}
%     % \vspace{-2em}
% \end{figure*}

In the first layer, all candidate's points are divided into $M$ different cluster buckets. Each cluster bucket contains $n_m$ data points and has a radius of $e_m$.
For the uniform distribution, $n_m = N/M$, and $e_1 = e_2 = \dots =e_m$. Therefore, the in-degree of all tokens in this layer are equal.   
% While for non-uniform distributions, $n_m $ and  $e_m$ exhibits significant variations.

In the second layer,  all input embedding is $\textbf{X}^{'}$, the residual of the first layer. Due to the difference in the magnitude of residual values, the input distribution in this layer is non-uniform. There are a large number of points with smaller magnitudes (points near the cluster centers in each bucket from the previous layer), which is equal to $n_m * M * \rho = N *\rho$, $\rho$ is the ratio. At the same time, there are small points with larger magnitudes, which are considered as outliers. To reduce the clustering loss, the clustering process in this layer focuses on these outliers. As a result, the points with smaller magnitudes will occupy fewer cluster centers, while the outliers will either occupy individual cluster centers or multiple cluster centers. Therefore, this layer's semantic IDs will form large routing nodes, exhibiting a long-tail phenomenon, which is also demonstrated in the second layer of Figure \ref{fig:Analysis}.

In the third layer, all input point magnitudes become consistent again and relatively uniform. Therefore, the code distribution in this layer is similar to the first layer, with a uniform distribution. As a result, it can be directly observed that the large routing nodes from the second layer diverge into multiple smaller nodes in the third layer, creating a one-to-many situation, as shown in the third layer of Figure \ref{fig:Analysis}. At the same time, if the residuals in the second layer tend towards zero, there will still be some clustering in the third layer. However, since all magnitudes are very small at this point, the impact of the clustering effect is limited.

As we continue to iterate through the layers, this phenomenon of non-uniform distribution and long-tail clustering followed by uniform distribution will alternate. However, as the number of layers increases, the residuals become smaller (refer to layer 4 of Figure \ref{fig:Analysis}), and the clustering effect weakens, so it can be ignored. Ultimately, this leads to the formation of an hourglass-like structure, where the input data is first compressed into a smaller number of clusters, then expands back out into a larger number of clusters, and finally converges to a uniform distribution. Upon the completion of SID construction, the influence of the RQ quantization method, coupled with the dominance of head tokens in the intermediate layer, naturally leads to the sparsity of paths.

\begin{table*}[tp]
    \centering
    \caption{The performance of generative retrieval on E-commerce datasets with RQ3x12, i.e., $L=3, M=2^{12}$. The head/tail token denotes the head/tail semantic ID in the second layer, respectively. }
    \begin{tabular}{c|cccccc}
    \toprule
        Method &  Recall@1 & Recall@3 & Recall@5 & Recall@10 & Recall@30 & Recall@50 \\
    \midrule
        LLaMA2-0.8B\textsuperscript{*}  & 0.2480  & 0.4080 & 0.4990 & 0.590 & 0.7080 & 0.7480 \\
            \textit{Head Token} & 0.3617 & 0.5745 & 0.6894 & 0.7745 & 0.8894 & 0.9191 \\
            \textit{Tail Token} & 0.2131 & 0.3569 & 0.4405 & 0.5333 & 0.6523 & 0.6954 \\
     \midrule
        Qwen1.5-7B & 0.2770 & 0.4720 & 0.5700 & 0.6600 & 0.7700 & 0.7930 \\
          \textit{Head Token} & 0.3450  & 0.5970  & 0.7040  & 0.8020  & 0.8960  & 0.9120  \\
        \textit{Tail Token} & 0.2470  & 0.4160  & 0.5100  & 0.5950  & 0.7190  & 0.7470  \\
    \midrule
        Baichuan2-7B & 0.2730   & 0.4900  & 0.5900  & 0.6760  & 0.7670  & 0.8040 \\
          \textit{Head Token} & 0.3440  & 0.6000  & 0.7200  & 0.8140  & 0.9020  & 0. 9210 \\
          \textit{Tail Token} & 0.2480  & 0.4360  & 0.5250  & 0.6110  & 0.7180  & 0.7540  \\
    \midrule 
          Given Layer 1\textsuperscript{*} & 0.340  & 0.497  & 0.567  & 0.632  & 0.722  & 0.756  \\
          Exchange Layer 1\&2\textsuperscript{*} & 0.2390  & 0.4190  & 0.5100  & 0.6070  & 0.7150  & 0.7540  \\
            + Given Layer 1\textsuperscript{*} & \textbf{0.6600} & \textbf{0.8240} & \textbf{0.8650} & \textbf{0.8910} & \textbf{0.9160} & \textbf{0.9190} \\
    \bottomrule
    \end{tabular}
    \label{base-method}
    \begin{tablenotes}
    % \begin{itemize}
        \item \textsuperscript{*} \small{These experiments are based on the LLaMA2-0.8B model, which adopts the LLaMA2 structure and SFT on Chinese corpora.}
    % \end{itemize}
    \end{tablenotes}
    % \vspace{-1em}
    
\end{table*}

Similarly, for the un-uniform distribution, such as long-tail distribution, the residual distribution becomes even more uneven, resulting in a more severe phenomenon.

\subsection{Impact on the GR}
% In the previous section, we analyzed that there is an obvious long-tail utility in the second layer of semantic ID, which means there are one-to-many and many-to-one situations. This phenomenon will seriously affect the generation of downstream tasks such as GR (Generative Retrieval).

In the above section, we have discussed the long-tail distribution in the second layer of Semantic ID, indicating a one-to-many and many-to-one structure. We argue that this phenomenon significantly impacts the generation of downstream tasks, especially for generative retrieval task.

To measure this impact, we conducted various experiments. First, we altered the distribution of Semantic ID by interacting with the first and second layers. On this basis, we only predicted the tokens of the second and third layers while keeping the tokens of the first layer fixed. 

During the evaluation process, we divide the test set into two groups according to the distribution of second-layer tokens: the head token test set and the tail token test set. As shown in Table \ref{base-method}, the performance of the head token test set significantly improved, whereas the performance of the tail token test set was notably poorer. This performance disparity can be attributed to the previously analyzed path sparsity and long-tail distribution of tokens, leading to biased results. This phenomenon has been observed across models of different scales (LLaMA2, Baichuan2, and Qwen1.5) and different parameters of RQ, highlighting the widespread impact of long-tail token distribution and path sparsity on model performance. 

To further investigate the impact of the hourglass phenomenon on model performance, we conduct three critical experiments: 1) give the first token directly as input, 2) exchange the tokens of the first and second layers, and 3) give the first token of the swapped sequence as input.

Swapping only the first and second layers results in a significant long-tail distribution in the first layer, and the issue of the long-tail distribution remains unresolved. As shown in Table \ref{base-method}, the changes in metrics are minimal.
However, if we swap the layers and provide the 1st token, the task shifts to predicting the 2nd and 3rd layer. This simplifies the task since the true first-layer is given, mitigating the long-tail distribution's impact and significantly improving performance. Conversely, if we don't swap the layers and still provide the first token, the second-layer SID maintains its long-tail distribution. These results shown in Table \ref{base-method} are higher than the baseline but worse than when the first token is given after swapping.

These approaches aim to mitigate the effects of the long-tail distribution, and results verify a significant improvement. This finding indicates that the hourglass phenomenon has a substantial negative impact on model performance. Through the above experiments, we not only confirmed the existence of the hourglass effect but also elucidated its specific impact on model performance, thereby providing a robust basis for future optimization.

\begin{table*}[t]
    \centering
    \caption{The performance of generative retrieval on E-commerce based on RQ3x12.}
    \resizebox{\textwidth}{!}{
    \begin{tabular}{c|cccccc}
    \toprule
        Method &  Recall@1 & Recall@3 & Recall@5 & Recall@10 & Recall@30 & Recall@50 \\
    \midrule
        LLaMA2-0.8B  & 0.2480  & 0.4080 & 0.4990 & 0.590 & 0.7080 & 0.7480 \\
       
     \midrule
         Focal Loss \cite{lin2017focal} & 0.2310    & 0.4270   & 0.5050   & 0.6110   & 0.7300   & 0.7640  \\
        Mile Loss \cite{su2024mile} & 0.2590  & 0.4380  & 0.5110  & 0.6090  & 0.7250  & 0.7600  \\
    \midrule
        Remove 2-th layer & 0.3090  & 0.4310  & 0.4970  & 0.5640  & 0.6580  & 0.7020  \\
        Remove 2-th layer top@20 & 0.2500   & 0.4270   & 0.5130   & 0.6120   & 0.7250   & 0.7580  \\
        Remove 2-th layer top@200 & 0.3190   & 0.4740   & 0.5600   & 0.6550   & 0.7450   & 0.7760  \\
        Remove 2-th layer top@400 & \textbf{0.3340}  & \underline{0.5070}  & \textbf{0.5950}  & \textbf{0.6800}  & \textbf{0.7760}  & \underline{0.7990}  \\
        Remove 2-th layer top@600 & \underline{0.3320}   & \textbf{0.5080}   & \underline{0.5850}   & \underline{0.6720}   & \underline{0.7700}   & \textbf{0.8010}  \\
   
    \bottomrule
    \end{tabular}
    }
    \label{remove top k}
    % \vspace{-1em}
\end{table*}

\section{Methods and Experiments}
To alleviate the hourglass effect, we propose two simple yet effective methods.
\subsection{Heuristic Method}
One heuristic approach is to directly remove the second layer, eliminating the impact of the long tail. However, it can lead to insufficient spatial capacity,  i.e., $M^L \rightarrow M^{L-1}$.
Note that, here needs first to generate an $L$-layer SID and then remove the second layer, which differs from directly generating a two-layer SID, where large routing nodes may still exist. 

% By removing the second layer after generating the full $L$ layer SID, we can effectively reduce the impact of the hourglass effect while maintaining some of the benefits of having multiple layers.

\subsection{Variable Length of SID}
Another simple method is to adaptively remove the top tokens of the second layer, making the semantic ID a variable-length structure. Here, a top@K strategy is used, with p as a threshold. This approach ensures that the distribution remains unchanged while reducing the impact of the hourglass effect selectively. What's more, the spatial capacity is sufficient, i.e., $M^L \rightarrow  M^L + K(M^{L-2} - M^{L-1})$. Note that the choice of top-k depends on the actual data distribution, so ablation testing is necessary. In summary, while this method is simple and efficient, it is not optimal and can only alleviate, but not completely resolve, the hourglass phenomenon.
 % K/M * M^{L-1} + (M-K)/M * M^L =

\subsection{Experiments}
To further validate the effectiveness of the method, experiments are conducted on the LLaMA model and on a real large-scale e-commerce platform. We randomly selected hundreds of millions of training samples from nearly sixty days of data, with a user base reaching tens of millions and a product catalog of two hundred million items \cite{li2023adaptive,wang2023learning}. The average length of user behavior sequences is 100.

Results indicate that by applying the adaptive token removal strategy, the performance of the model is improved while maintaining a similar computational cost compared to the base model, and several objective optimizations, such as Focal Loss \cite{lin2017focal} and Mile Loss \cite{su2024mile}.

Specifically,  experimental results showed that the model with top@400 token removal outperforms the baseline model in terms of most evaluation metrics. This suggests that the method effectively reduces the impact of the long-tail effect. As the number of tokens removed increases, the performance improvement of the model encounters a bottleneck. Especially when all tokens are removed, this limitation is particularly pronounced, which is presumed to be due to the absence of long-tail tokens, resulting in a loss of recall. At the same time, removing the second layer directly will cause one SID to correspond to multiple items.  

This fine-grained analysis provides strong evidence for the effectiveness of the proposed method, which selectively removes less important tokens while retaining the most informative ones, leading to improved model performance even when a substantial amount of data is removed.

\subsection{Valid Ratio}
\begin{figure}
    \centering
    \includegraphics[width=1\linewidth]{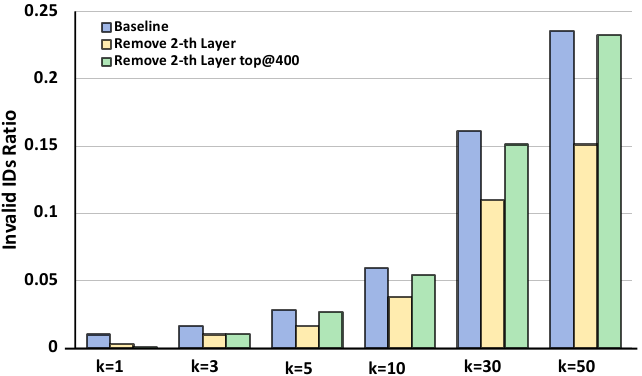}
    \caption{Invalid IDs Ratio when generating Semantic IDs using Beam Search for various values of $k$}
    \label{fig:InvalidRatio}
    % \vspace{-0.5cm}
\end{figure}

During the autoregressive decoding process, as the model decodes the next token of the target SID, it may predict invalid SIDs, SIDs that are not in the SID's vocabulary, or do not correspond to any item in the full dataset. Therefore, we have calculated the proportion of invalid SIDs on the LLaMA2-0.8B model with RQ3x12.  As shown in Figure \ref{fig:InvalidRatio},  we can see the base model, the invalid ratio of the proposed method is lower than the base model, indicating that the higher-quality generation items with a lower ratio of hallucination. Furthermore, when the number of recalls is less than 10, the invalid ratio is below 5\%. Thus, the effectiveness of generation is to meet practical needs. In other situations, where a higher number of recalls is required (k=50), the invalid ratio is higher. Across various sizes of base models and different RQ parameter settings, the results tend to converge on the same conclusion. Therefore, it is necessary to employ the retrieval augmented generation (RAG) \cite{lewis2020retrieval,ding2024retrieve} for processing during the inference process, such as prefix-tree \cite{beurer2024guiding}, and FM\_Index \cite{herruzo2021enabling}.

\section{Conclusion}
This study systematically explores the limitations of RQ-SID in GR, particularly identifying the "hourglass" phenomenon in the intermediate layer where codebook tokens are overly concentrated, leading to path sparsity and long-tail distribution. Through extensive experiments and ablation studies, we have demonstrated the existence of this phenomenon and conducted an in-depth analysis attributing its root cause to the characteristics of residuals. To alleviate this issue, we propose two methods: a heuristic approach that removes the second layer and a variable-length token strategy that adaptively adjusts token distribution. Experimental results show both methods effectively mitigate the bottleneck effect, with the adaptive token distribution adjustment yielding the best results. While this method is simple and efficient, it is not optimal and can only alleviate, but not completely resolve, the hourglass phenomenon. To the best of our knowledge, this is the first systematic exploration of the deficiencies of RQ-SID in GR, providing a solid foundation for future model optimizations and significantly enhancing model performance by improving codebook utilization. 

\vspace{15em}

\section*{Acknowledgments}
This work has been supported by the National Key R\&D Program of China under grant No. 2022YFF0902500, the National Natural Science Foundation of China under grant No.62472447, the Science and Technology Innovation Program of Hunan Province under grant No.2023RC1023, the Hunan Provincial Natural Science Foundation of China under grant No.2024JK2006, and Shenzhen Startup Funding No. QD2023014C.

\bibliography{custom}
% \clearpage
% \newpage

\appendix
% \vspace{25em}
\section{Distribution and Connections of Different RQ-Semantic IDs}
\begin{figure}[h]
    \includegraphics[width=1\linewidth]{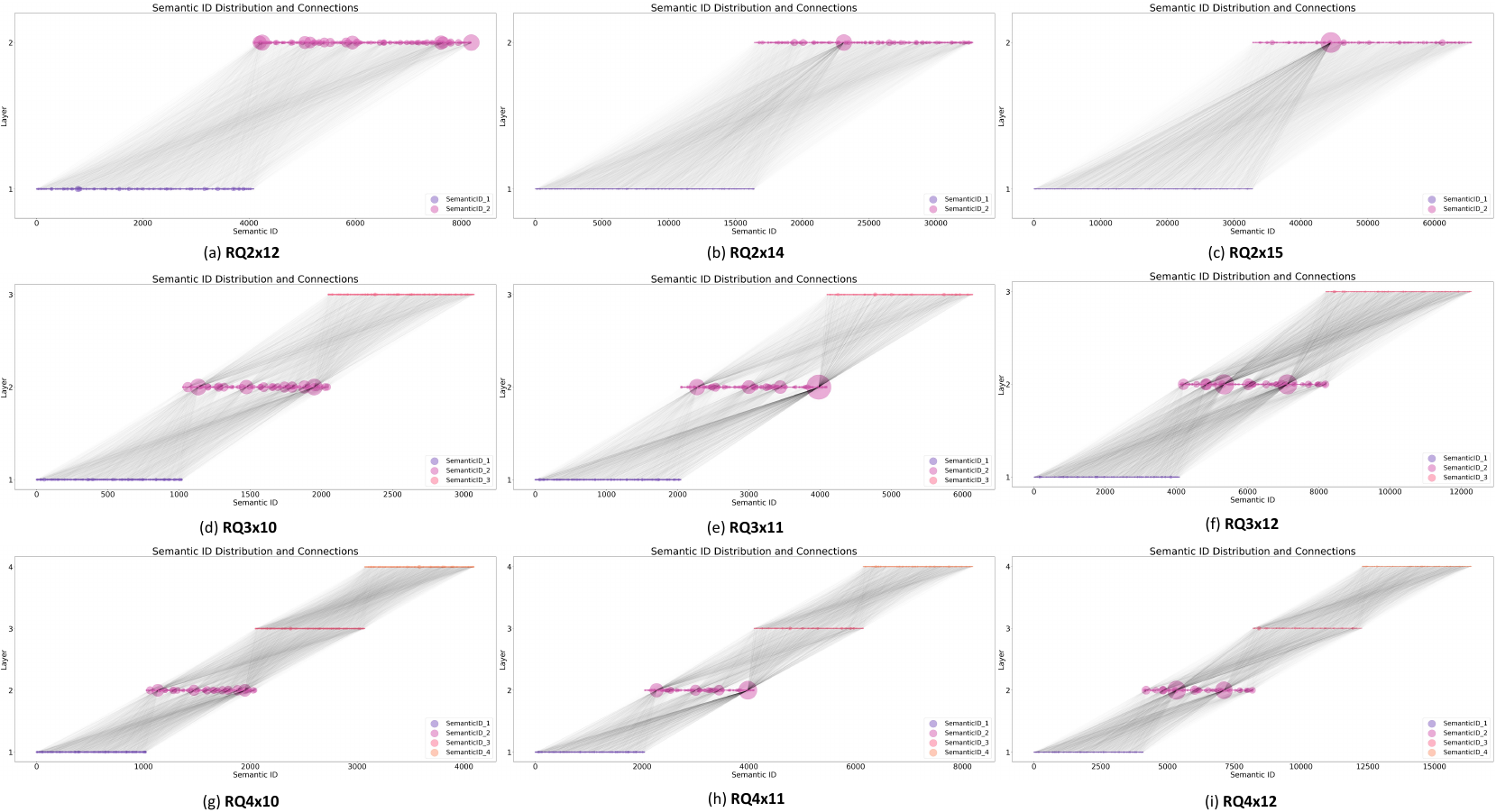}
    \caption{Distribution and Connections of Different RQ-Semantic IDs}
    \label{fig:dis-label}
\end{figure}
\balance
% This is an appendix.

\end{document}